\documentstyle[12pt]{article}
\newcommand{\al}{\alpha}
\newcommand{\olocl}{\int_{0}^{1}dt\int_{0}^{1}d\al}
\newcommand{\usual}{{4Q^2 \over M_W^2}t^2 \al (1-\al)+i\epsilon}
\newcommand{\use}{{4Q^2 \over M_W^2}t^2 \al (1-\al)}
\newcommand{\usualp}{{4Q^2 \over M_W^2}t^3(3t-2) \al (1-\al)}
\begin{document}
\begin{titlepage}
\begin{centering}
{\Large{\bf Trilinear Gauge Boson Vertices in the MSSM}}\\
\vspace{1.5cm}
{{\bf A. B. Lahanas }\hspace{.02cm}}$^{\dag}${\hspace{.1cm}} and
{\hspace{.1cm}}{{\bf V. C. Spanos }\hspace{.02cm}}$^{\ddag}$  \\
\vspace{.4cm}
University of Athens, Physics Department,
Nuclear and Particle Physics Section,\\
\vspace{.05in}
Ilissia, GR -- 15771  Athens, GREECE\\
\vspace{1.2cm}
{\bf Abstract}\\
\vspace{.1in}
\end{centering}
{
\noindent
{{We study the $C$ and $P$ even $WW\gamma$ and $WWZ$ trilinear gauge boson
vertices (TGV's),
in the context of the MSSM as functions of the soft SUSY breaking parameters
$A_0, m_0, M_{1/2}$ and the momentum $q$ carried by  $\gamma, Z$, assuming the
external $W$'s are on their mass shell. We follow a complete renormalization
group analysis taking into account all constraints imposed by the radiative
breaking of the electroweak symmetry. It is found that for energies
${\sqrt{s}} \equiv q^2 \le 200 \; GeV$  squark and slepton contributions to
the aforementioned couplings are two orders of magnitude smaller than those
of the Standard Model (SM). In the same energy range the bulk of the
supersymmetric Higgs corrections to the TGV's is  due to the lightest neutral
Higgs, $h_0$, whose contribution is like that of a Standard Model Higgs
of the same mass.
The rest have negligible effect due to their heaviness. The
contributions of the Neutralinos and Charginos are sensitive to the input
value for the soft gaugino mass $M_{1/2}$ being more pronounced for values
$M_{1/2} < 100 \; GeV$. In this case and in the unphysical region,
$0 < \sqrt{s} < 2 M_W $ their contributions are substantially enhanced
resulting to large corrections to the static quantities of the $W$ boson.
However such an enhancement is not observed in the physical region and their
corrections to the TGV's are rather small. In general for
$2 M_W < \sqrt{s} < 200 \; GeV $ the MSSM predictions differ from those of the
SM but they are of the same order of magnitude. Deviations from the SM
predictions to be detectable require sensitivities reaching the per mille
level and hence unlikely to be observed at LEP200. For higher energies  SM and
MSSM predictions exhibit a fast fall off behaviour, in accord with unitarity
requirements, getting smaller by almost an order of magnitude already at
energies $\sqrt{s} \approx .5 \; TeV $. At these energies the task
of observing deviations from the SM predictions, which are due to
supersymmetry, becomes even harder requiring higher accuracies.}}}
\paragraph{}
\par
\vspace{.5cm}
\begin{flushleft}
Athens University\\
UA/NPPS-18,\quad  April 1995\\
\end{flushleft}
\rule[0.in]{4.5in}{.01in}\\
E-mail:
$^{{\hspace{.2cm}}\dag}${\hspace{.02cm}
alahanas@atlas.uoa.ariadne-t.gr} ,\quad
$^{{\hspace{.2cm}}\ddag}$ { \hspace{.02cm}
vspanos@atlas.uoa.ariadne-t.gr}
\end{titlepage}
%
\section*{1. Introduction}
The Standard Model (SM) as been remarkably succeful in describing particles
interactions at energies around the $\sim 100\:GeV$. Precise measurements at
LEP provided accurate tests of the standard theory of electroweak
interactions $^{\cite{hollik,schwarz}}$ but we are still lacking a
direct experimental confirmation
of the non-abelian structure of the standard theory. The $WW\gamma$, $WWZ$ and
$ZZ\gamma$ couplings are uniqely determined within the context of the SM and
such couplings will be probed in the near future with high accuracy. The study
of the trilinear gauge bosons vertices (TGV's) in the
$e^{+}e^{-}\rightarrow W^{+}W^{-}$ process is the primary motivation for the
upgrading of LEP200 $^{\cite{treille}}$ and the potential for measuring
these has been discussed
in detail $^{\cite{treille,bil}}$. At an energy of about $190\:GeV$ and with
integrated luminosity
of $500\:pb^{-1}$ an accuracy of $.1$ for the determination of these
couplings
can be obtained. So far there are no stringent experimental bounds on these
couplings $^{\cite{ua2}}$ and the efforts of the various experimental groups
towards this direction are still continuing. In the near or remote future with
the proposed or already under construction high energy colliders (LHC, NLC,
CLIC, JLC) further improvements on the TGV's bounds will be obtained reaching
accuracies ${\cal O}(10^{-2}-10^{-3})$ $^{\cite{flei}}$. Such a precise
measurements are of vital importance not only for the SM itself but also for
probing new physics which opens at scales larger than the Fermi scale.
\par
The gauge boson vertex has been the subject of an intense theoretical study
the last years. In particular the $WWV$ vertex $(V=\gamma$ or $Z)$ has been
analysed in detail within the framework of the standard theory, as well as
in extension of it, and its phenomenology has been discussed. The lagrangian
density describing the $WWV$ interaction is given by $^{\cite{gounar,falk}}$
\begin{eqnarray}
& {\cal L}^{WWV}=-ig_{WWV}[(W_{\mu\nu}^{\dagger}W^{\mu}V^{\nu}-h.c.)+
                 {\kappa}_{V}W_{\mu}^{\dagger}W_{\nu}F^{\mu\nu}+
                 \frac{{\lambda}_{V}}{M_{W}^{2}}W_{\lambda\mu}^{\dagger}
                 W^{\mu}_{\nu}
                 V^{\nu\lambda}+...\:],\\
& g_{WWV}=\left\{ \begin{array}{ll}
                   e & \mbox{for $V=\gamma$} \nonumber \\
     e\cot\theta_{W} & \mbox{for $V=Z$} \nonumber
                \end{array}
        \right.
\end{eqnarray}
where the ellipsis stand for $P$ or $C$ odd terms and higher
dimensional operators.
In Eq. (1) the scalar
components for all gauge bosons involved have been omitted, that is
${\partial}{\cdot}W={\partial}{\cdot}V=0$, since essentially they
couple to massless fermions
\footnote{For on shell $W$'s we have
$({\Box}+m_{W}^{2})W_{\mu}=0$ and certainly ${\partial}{\cdot}W=0$,
while for the photon ${\partial}{\cdot}A$ vanishes on account of current
conservation.}.
At the tree level  $\kappa_{V}$ and $\lambda_{V}$
have the values $\kappa_{V}=1$, $\lambda_{V}=0$. However radiative
corrections modify these, the order of magnitude of these corrections
being ${\cal O}(\frac{\alpha}{\pi})\sim10^{-3}$.
Sensitivity limits of this order of magnitude
will not be reached at LEP200 but can be
achieved in future colliders where the TGV's can be studied in detail and
yield valuable information not only for the self consistency of the SM
but also for probing underlying new physics. Any new dynamics whose onset
lies in the $TeV$ range modifies $\kappa_{V}$, $\lambda_{V}$ and
deviations from the SM predictions are expected.

Supersymmetry (SUSY), is
an extension of the SM which is theoretical motivated but without any
experimental confirmation. The only experimental hint for its existence
derives from the fact that the gauge couplings unify at energies
$\sim10^{16}\:GeV$ if we adopt a supersymmetric extension of the SM in which
SUSY is broken at energies
$M_{SUSY}{\sim}{\cal O}(1)\:TeV$ $^{\cite{amaldi}}$ .
Supersymmetric particles with such large masses can be produced in the
laboratory provided we have very high energies and luminocities.
However their existence affects
$\kappa_{V}$ and $\lambda_{V}$, even at
energies lower than the SUSY production threshold making them deviate from
the SM predictions. Therefore the study of these quantities may
furnish as a good
laboratory to look for signal of supersymmetry at energies below the SUSY
production threshold. In any case such studies serve as a complementary
test along with other efforts towards searching for signals of new physics and
supersymmetry is among the prominent  candidates.
\par
In the SM $\kappa_{V}$ and
$\lambda_{V}$ as functions of the momentun $q^{2}$ carried by the $V$
boson $(V=\gamma,Z)$, for on shell $W$'s, have been studied in detail but
a similar analysis has not been carried out within the context of the MSSM.
Only the quantities $\kappa_{\gamma}(q^{2}=0)$, $\lambda_{\gamma}(q^{2}=0)$
have been considered which are actually related to the static quantities
magnetic dipole $(\mu_{W})$ and electric quadrupole $(Q_{W})$ moments
of the $W$ boson. To be of relevance for future collider experiments
the form factors $\kappa_{\gamma,Z}$, $\lambda_{\gamma,Z}$ should be
evaluated in the region $q^{2}>4M_{W}^{2}$. The behavior of
$\kappa_{\gamma,Z}$, $\lambda_{\gamma,Z}$ in this physical region may
be different from that at $q^{2}=0$  especially when the energy
gets closer to $M_{SUSY}$  and supersymmetric particles may yield sizable
effects, due to the fact that we are approaching their thresholds.
In those cases an
enhancement of their corresponding contributions is expected, unlike
SM  contributions which in this high energy regime are suppressed.
We should also point out that some of the supersymmetric
particles may have relatively small masses, for a certain range of
the parameters and in those circumstances their  contributions to TGV's
are not necessarily small. In order to know the magnitude of these effects a
detailed computation of the trilinear gauge boson couplings should be carried
out.
\par
In this work we undertake this problem and study the $C$ and $P$ even
 $WW\gamma$, $WWZ$ vertices in the context of the MSSM when the external
$W$ bosons are on their mass shell. Such studies are important in
view of forthcoming experiments at LEP200 and other future collider
experiments which will probe the structure of the gauge boson couplings and
test with high accuracy the predictions of the SM. If deviations from the SM
predictions are observed these experiments will signal the presence of new
underlying dynamics which opens at scales larger than the Fermi scale.
\par
The magnitude of the aforementioned couplings and their dependence on the
arbitrary parameters of the MSSM requires a systematic study in which all
limitations imposed by a renormalization group analysis of all running
parameters involved and especially those arising from the radiative breaking
of the electroweak (EW) symmetry are duly taken into account. In this paper
we deal with this issue and calculate the trilinear vector boson couplings as
functions of the momentum $q^2$ carried by the $\gamma,Z$ and the arbitrary
parameters of minimal supersymmetry.\\ \\
\newpage
This paper is organized as follows:\\
In section 2 we give a brief outline of the MSSM. In section 3 we carry on
to discuss the SM predictions for the TGV's paying special attention to
the contributions of fermions and those of the gauge bosons discussing
the issue of gauge independence. In section 4 the MSSM predictions for
these couplings is discussed. Section 5 deals with the absortive (Imaginary)
parts of these vertices and in section 6 we present the  numerical analysis
and discuss our conclusions.

\vspace{1cm}
\section*{2. The MSSM}
The MSSM is described by a Lagrangian $^{\cite{Nilles}}$
\begin{equation}
{\cal L}={\cal L}_{{SUSY}}+{\cal L}_{{soft}}
\end{equation}
where its  supersymmetric part ${\cal L}_{{SUSY}}$ is derived from
a superpotential ${\cal W}$ bearing the form
\begin{equation}
{\cal W}=(h_U \hat{Q}^i \hat{H}_2^j \hat{U}^c
         + h_D \hat{Q}^i \hat{H}_1^j \hat{D}^c
         + h_E \hat{L}^i \hat{H}_1^j \hat{E}^c
         + \mu \hat{H}_1^i \hat{H}_2^j) \epsilon_{ij} \quad,\quad
          \epsilon_{12}=+1
\end{equation}
In Eq.(3)  carrets denote supermultiplets. Minimality is enforced by assuming
the $SU(3)\times SU(2)\times U(1)$ as the gauge group and the least
number of chiral multiplets necessary to accomodate  matter fermions
and drive EW symmetry breaking. The superpotential above
conserves
$R$-parity. The $H_{1}$, $H_{2}$ Higgses give mass to up and down fermions
respectively after EW breaking takes place.
The part responsible for the soft breaking of supersymmetry
is given by
\begin{eqnarray}
- {\cal L}_{{soft}}&=&{\sum_{i}} m_i^2 |\Phi_i|^2
         + (h_U A_U Q H_2 U^c
         + h_D A_D Q H_1 D^c
         + h_E A_L L H_1 E^c+h.c.)\nonumber\\
                         &+&(  \mu B H_1 H_2 +h.c.)
         +\frac {1}{2} \sum_a M_a {\bar{\lambda}}_a \lambda_a .
\end{eqnarray}
where $m_{i}$ are the soft scalar masses, $A_{U,D,L}$ are trilinear soft
couplings, $B$ is the Higgs mixing parameter and $M_{\alpha}\:,\:\alpha=1,2,3$
are the soft gauginos masses for the $U(1)$, $SU(2)$ and $SU(3)$ gauge
fermions
respectively. Throughtout this paper we assume universal boundary conditions
at the unification scale $M_{GUT}\simeq10^{16}\:GeV$
\begin{equation}
m_{i}^{2}\equiv m_{0}^{2} \quad, \quad A_{U,D,L}\equiv A_{0}\quad,
\quad M_{\alpha}\equiv
M_{1/2}
\end{equation}
This choice is suggested by grand unification  and also by absence of
flavor changing neutral currents
which puts stringent constraints on the difference of masses squared  of the
same charge squarks. However this is in no way mandatory.
At least it parametrizes our ignorance concerning the origin
of the soft SUSY breaking parameters in the most economical and plausible
way. We don't expect that altering the boundary
conditions given in Eqs.(5) above will drastically affect the
estimates for the TGV's.
\par
The values of all running parameters
in the vicinity of the electroweak scale are then given by solving their
Renormalization Group Equations (RGE's)
having as initial conditions Eqs. (5) $^{\cite{Iban}}$.
\par
The breaking of the EW symmetry is known to proceed via radiative
corrections driven by the large top Yukawa coupling $^{\cite{Iban}}$.
The equations minimizing the scalar potential of the theory are
\begin{equation}
\frac {M_{ Z}^{ 2}}{2}=\frac {\bar {m}^{ 2}_{ 1}-
   \bar {m}^{ 2}_{ 2}  tan^{ 2}\beta}
   {tan^{ 2}\beta- 1}\quad,\quad
                         sin2\beta=-\frac {2B\mu}
                        {\bar {m}^{ 2}_{ 1}
                        +\bar {m}^{ 2}_{ 2}}
\end{equation}
where the angle $\beta$ sets the relative strength of the v.e.v. of the
$H_{1}$ and $H_{2}$ Higgs fields involved
\begin{equation}
\tan\beta(M_{Z})\equiv \frac{v_{2}(M_{Z})}{v_{1}(M_{Z})}.
\end{equation}
In Eqs. (6)--(7) all quantities are meant at $M_{Z}$, the experimental value
of the $Z$ boson
mass $(M_{Z}=92.18\:GeV)$, and ${\bar m}_{1}^{2}\:,\:{\bar m}_{2}^{2}$ are
defined as
\begin{equation}
\bar {m}^{ 2}_{ {1,2}}={m}^{ 2}_{ {1,2}}+
\frac {\partial \Delta V}{\partial v_{ {1,2}}^{ 2}} \quad,
\quad  {m}^{ 2}_{ {1,2}}={m}^{ 2}_{H_{ {1,2}}}+
\mu^{ 2}\:.
\end{equation}
In Eq.(8) $m_{H_{1,2}}$ are the soft Higgses masses and ${\bar m}_{1,2}$
differ from $m_{1,2}$ by the contributions of the one loop corrections
$\Delta V$ to the
scalar potential of the theory. Including the one loop corrections within
the minimizing Eqs.(6), as prescribed by Eq.(8), is a necesssary ingredient
for the numerical stability of our physical results. If not included the
physical quantities would strongly  depend on the choice of the scale
at which physical quantities are evaluated leading to results that
are ambiguous and untrustworthy $^{\cite{castano}}$.
\par
The arbitrary parameters of the model are the soft parameter $m_{0}$,
$A$, $M_{1/2}$, $B$ the mixing parameter $\mu$ as well as the values
of the top Yukawa coupling $h_{t}$ {\footnote {The top Yukawa coupling
although localized in a narrow range of values in view of the recent
experimental evidence for the top quark $^{\cite{abe}}$ is being considered
as an input parameter since the occurrence of symmetry breaking is very
sensitive to its value.}}.
This number is reduced to
five if use is made of the first of the minimizing equations (6). Then a
convenient set of independent parameters, which is adopted by many authors,
is to take the set
\begin{equation}
m_{0} \quad,\quad A_{0} \quad,\quad M_{1/2} \quad,\quad
\tan\beta(M_{Z}) \quad,\quad m_{t}(M_{Z})
\end{equation}
where $m_{t}(M_{Z})$ is the value of the ``runing'' top quark mass at the
scale $M_{Z}$. This facilitates the numerical analysis a great deal since
the RGE's of all soft masses and parameters involved, with the
exception of  $B$ and $\mu$, do not
depend on $B$, $\mu$  (nearly decouple). Given the inputs (9)
the RGE's can be solved and predictions for the mass spectrum can be given.
At this point we should remark that all subtleties associated with this
approach, like for instance the presence of low energy thresholds and other
uncertainties due to higher loop effects, in no way affect the one loop
corrections to the TGV's.
Complete expressions for the RGE's of all
parameters involved can be traced in the literature and will not be repeated
here $^{\cite{Nilles,Iban}}$.
\par
After this brief outline of the MSSM we embark
to discuss the TGV's defined in the previous section.
\vspace{1cm}
\section*{3. The SM contribution to
$\Delta k_{V}(Q^{2})$, $\Delta Q_{V}(Q^{2})$}
Although the SM contributions to the TGV's have already been calculated
in the literature $^{\cite{argy}}$, for reasons of completeness
we shall briefly discuss them in this section too  paying special
attention to the contributions of fermions and gauge bosons.
\par
In momentum space the most general $WWV$ vertex $(V=\gamma$ or $Z)$ with
the two $W$'s on shell and keeping only the transvers degrees of freedom for
the $\gamma$ or $Z$ can be writen as $^{\cite{gounar}}$
\begin{eqnarray}
\Gamma_{\mu \alpha \beta}^{V}& = &-ig_{WWV} \, \{ \,
                     f_{V}[2g_{\alpha \beta}{\Delta_\mu}
       +4(g_{\alpha \mu}{Q_\beta}-g_{\beta \mu}{Q_\alpha})] \nonumber\\
   &   &+2 \, {\Delta k}_V \,
                          (g_{\alpha \mu}{Q_\beta}-g_{\beta \mu}{Q_\alpha})+
       4 \, {\frac  {{\Delta Q}_V}  {M_{ W}^{ 2}}} \, {\Delta_\mu}
       ({Q_\alpha}{Q_\beta}-{\frac {Q^{ 2}}{2}} g_{\alpha \beta})\}+...
\end{eqnarray}
where ${{\Delta k}_V} \, \equiv \, k_V + {\lambda}_V -1 \, , \,
{{\Delta Q}_V} \, \equiv \, -2 \,  {\lambda}_V $.
The kinematics of the vertex is shown in  Fig. 1 .
\par
The ordinary matter fermion
contributions both to $Q^{2}=0$ and $Q^{2}\neq 0$  have been studied
elsewhere $^{\cite{bard,couture}}$. However in those works there is an
important sign error which affects substantially the results given
in those references $^{\cite{spa}}$. This has been also pointed out
independently in ref. ${\cite{cula}}$ .
The  consequences of this  for the static quantities of the $W$
boson $\mu_{W}$, $Q_{W}$ has been discussed in detail in ref.
${\cite{spa,lahanas}}$.
In the  massless fermion limit,
which is actually the case for the first two families, this leads to
nonvanishing contributions for both $\Delta k_{\gamma}(Q^{2}=0)$ and
$\Delta Q_{\gamma}(Q^{2}=0)$, contrary to what had been previously claimed.
These are proportional to $Tr(QT_{3})\neq 0$,
unlike the anomaly terms which are proportional to $TrQ$
and hence vanishing. The details of this calculation which  points out this
important sign error can be traced in the literature $^{\cite{spa}}$.
In the SM these contributions to $\Delta k_{\gamma}(Q^{2}=0)$ are large and
negative partially cancelling the contributions of the gauge bosons and
the standard model Higgs which are positive.
\par
In units of $g^2/{16\pi^{2}}$
the fermion contributions of the triangle graph shown in Fig. 2a
are as follows,
%
\begin{eqnarray}
\Delta k_{V}&=&-c_{V} T_3^f  C_g    \olocl  \{g_{L}^{f}
     [t^4+t^3(-1+R_{f'}-R_f)+t^2(R_f-R_{f'}) \nonumber\\
    & &+\frac{4Q^2}{M_W^2}t^3(3t-2)\alpha(1-\alpha)]
          +g_R^f [\, R_f \, {t^2}]  \} {1\over L_f^2}      \\
\Delta Q_{V}&=& -c_{V} T_3^f  C_g \, g_{L}^{f}\int_{0}^{1}dt
    \int_{0}^{1}d\alpha {8t^3(1-t)(1-\alpha)\alpha \over L_f^2}
\end{eqnarray}
where $c_{\gamma}=1,\, c_Z=R$ with  $R\equiv M_Z^2/M_W^2$.
In Eqs. (11) and (12)
\begin{eqnarray}
L_f^2 \equiv  t^2(1-{4Q^2 \over M_W^2}\alpha(1-\alpha))+t(R_f-R_{f'}-1)
+R_{f'}+i\epsilon \quad , \quad R_{f,f'} \equiv {m_{f,f'}^2 \over M_W^2}
\nonumber\
\end{eqnarray}
and $C_g$ is the color factor (1 for leptons, 3 for quarks). The couplings
$g_{L,R}^f$ appearing in Eq.(11) are as follows
\begin{eqnarray}
V=\gamma \quad & : & \quad\quad g_L^f=g_R^f=Q^f_{em} \nonumber\\
V=Z      \quad & : & \quad\quad g_L^f=Q^f_{wL} \, , \,
                g_R^f=Q^f_{wR}  \nonumber\
\end{eqnarray}
where $Q^f_{em}$ are the electromagnetic charges, and $Q^f_{wL,R}$ are the
weak charges for left/right handed fermions, which are defined by the relation
$Q^f_w\equiv T_3^f-Q^f_{em}\sin ^2 \theta _W$.
$T_3^f$ is the weak isospin of the fermion $f$, ie.
$T_3^f=-1/2$ for the left handed charged leptons etc. At zero momentum
transfer
$\Delta k_\gamma$, $\Delta Q_\gamma$ are nonvanishing as said earlier
yielding ${g^2 \over {16\pi^2}}{\cal O}(1)$ contributions to both
$\Delta k_\gamma$, $\Delta Q_\gamma$. Actually it turns out that when $Q^2=0$
this is the larger contribution to $\Delta Q_\gamma$ of all sectors, while
$\Delta k_\gamma (Q^2=0)$ is negative partially cancelling large
$(\sim {g^2 \over {16\pi^2}}{\cal O}(1))$ contribution from gauge boson and
Higgs particles as said previously.
We should point out that $\Delta k_V$, $\Delta Q_V$ as
given in Eqs.(11) and (12) refer to both the real and imaginary parts
of the vertices;
as we pass the internal particle threshold imaginary parts develop, too due
to the $i\epsilon$ appearing in the denominator. One can easily check that
as $Q^2\rightarrow \infty$  fermion contributions to both
$\Delta k_V$, $\Delta Q_V$, tend to zero as demanded by unitarity.
\par
Regarding the contributions of the gauge bosons to $\Delta k_V$, $\Delta Q_V$
the calculations were carried out in the 't Hooft -- Feynman gauge
$^{\cite{argy}}$, and the results are known to be gauge dependent.
The details of this calculation can be traced in the literature
[see Eq.(8)--(23) of ref. ${\cite{argy}}$].
In order to render the trilinear gauge boson vertices gauge independent
we should add to them additional contributions from box graphs
by applying special field theory techniques, such as the
pinch technique $^{\cite{papa}}$, or work in manifestly gauge invariant
gauges $^{\cite{drenner}}$.
As a result of this gauge dependence the quantity $\Delta k_V$ turns out to
have bad high energy  behaviour, growing logarithmically as the energy
increases, violating unitarity constraints. Besides being gauge dependent
$\Delta k_V$ is also singular at the infrared ($IR$). Actually this $IR$
singularity occuring in one of the graphs is the only one surviving among
several other which cancel against each other. As said previously for
the restoration of gauge independence additional contributions from some box
graphs, the pinch contributions, should be appended to the vertex parts.
These also cancel the $IR$ divergence mentioned earlier.
In units of $g^2/16 \pi^2$ these pinch parts are given by $^{\cite{papa}}$,
\newpage
\begin{eqnarray}
{\Delta {\hat k}}_{\gamma}&=&-2 {Q^2 \over\ M_W^2} \{ \cos^2\theta_W
                                                            \int_{0}^{1}dt
    \int_{0}^{1}d\alpha {t^2-2t \over t^2+R(1-t)-\frac{4Q^2}{M_W^2}t^2
     \alpha(1-\alpha)+i\epsilon}\nonumber\\
  & &+ \sin^2\theta_W \int_{0}^{1}dt
     {1 \over 1-\frac{4Q^2}{M_W^2}t(1-t)+i\epsilon} \} +IR
\end{eqnarray}
\begin{eqnarray}
{\Delta {\hat k}}_{Z}&=&-{1 \over 2}({4Q^2 \over\ M_W^2}-R) \{
                                                            \cos^2\theta_W
    \int_{0}^{1}dt\int_{0}^{1}d\alpha {t^2-2t \over t^2+R(1-t)-
    \frac{4Q^2}{M_W^2}t^2 \alpha(1-\alpha)+i\epsilon}\nonumber\\
  & &+ \sin^2\theta_W \int_{0}^{1}dt
     {1 \over 1-\frac{4Q^2}{M_W^2}t(1-t)+i\epsilon}  \} +IR  \quad .
\end{eqnarray}
In Eqs. (13)--(14), $R\equiv M_Z^2/M_W^2$ while $IR$ stands for the infrared
singularities which cancel against the corresponding singularities of the
vertices given in ref. ${\cite{argy}}$. Once the pinch contribution
Eq.(13)--(14) are taken into account the gauge boson contributions
become gauge independent
approaching zero values as $Q^2$ increases as demanded by unitarity and are
also free of infrared singularities.
\vspace{1cm}
\section*{4. The MSSM contribution to
$\Delta k_{V}(Q^{2})$, $\Delta Q_{V}(Q^{2})$}
At $Q^2=0$ the MSSM contributions to $\Delta k_{\gamma}$, $\Delta Q_{\gamma}$
have been studied elsewhere and their dependences on the soft breaking
parameters $A$, $m_0$, $M_{1/2}$, $\tan \beta$ and top quark mass $m_t$
have been investigated $^{\cite{spa}}$. The $\tilde{q}$, $\tilde{l}$
(squarks, sleptons), $\tilde{Z}$, $\tilde{C}$
(neutralinos, charginos) as well as the supersymmetric Higgs contributions
to $\Delta k_{V}$, $\Delta Q_{V}$ are deduced from the triangle
graphs shown in Figures 2 and 3. We show only graphs
that yield nonvanishing contributions to at least one of the
$\Delta k_{V}, \Delta Q_{V}$. In the following we shall consider
the contributions of each sector separately.
\subsection*{Squarks--Sleptons ($\tilde{q}$, $\tilde{l}$)}
\par
We first consider the contributions of the sfermion sector of the theory
which can be read from the diagram (b) of Figure 2. Unlike matter fermions
this graph involves mixing matrices due to the fact that left  ${\tilde f}_L$
and right  ${\tilde f}_L^c$  handed sfermions mix when electroweak symmetry
breaks down. Such mixings are substantial in the stops, due to
the heaviness of the top quark, resulting to large mass splitting of the
corresponding mass eigenstates ${\tilde t}_{1,2}$. For the sfermions we find
after a straightforward calculation that:
\newpage
\begin{eqnarray}
&& \Delta k_V=
       -2 C_g c_V T_3^{ f} g_L^f \sum_{i,j=1}^{2}
      {  ( {K_{i1}^{{\tilde f'}} }  {K_{j1}^{\tilde f})} }^2
   \olocl { t^2(1-t)(2t-1+R_{{\tilde f}_j}- R_{{\tilde f}^\prime_i}) \over
            L_{\tilde f  }^2} \nonumber  \\
&&                     \\
&& \Delta Q_V=
      2  C_g c_V T_3^{ f} g_L^f \sum_{i,j=1}^{2}
    {  ( {K_{i1}^{{\tilde f'}} }  {K_{j1}^{\tilde f})} }^2
  \olocl { {4 t^3(1-t)\al (1-\al)}  \over
      L_{\tilde f }^2 }       \\
&&               \nonumber     \\
\vspace{.5cm}
&&  L_{\tilde f}^2\equiv t^2+(R_{\tilde f_{j}}-R_{\tilde f^{\prime }_{i}}
        -1)t+R_{\tilde f^{\prime }_{i}} -\usual  \quad,\quad
  R_{ {\tilde f}_i , {{\tilde  f}^\prime}_i }
     \equiv {(m_ { {\tilde f}_i , {{\tilde  f}^\prime}_i }  /M_W)}^2
             \nonumber
\end{eqnarray}
The prefactors appearing in the integrals above $c_V, T_3^{f}$, $g_L^f$ and
$C_g$,
are exactly those of fermions, see Eqs. (11)--(12), since fermions and
their superpartners
carry same quantum numbers. ${\tilde f}_{1,2}$ and ${\tilde f'}_{1,2}$
denote the mass eigenstates while ${\bf K}^{{\tilde f},{\tilde f'}}$
diagonalize
the corresponding mass matrices, i.e.
${\bf K}^{{\tilde f},{\tilde f'}} {\bf {\cal M}}^{ 2}_{{\tilde f},{\tilde f'}}
{{\bf K}^{{\tilde f},{\tilde f'}}}^{\bf T} =diagonal$.
In the stop sector for instance, where such mixings are large, the
corresponding mass matrix is given by
\begin{equation}
{\bf {\cal M}}^{ 2}_{\tilde t}\,=\,\left(\begin{array}{lllll}
m^{ 2}_{ {Q_{ 3}}}+m^{ 2}_{ t}
&   &    &    & m_{ t}(A+\mu cot\beta)
\\
\quad\quad+{M^{ 2}_{ Z}}(cos2\beta)({\frac {1}{2}}
                       -{\frac {2}{3}}sin^{ 2}{\theta_{ W}}) &  &  &   &\\
 &  &  &  &  \\
 m_{ t}(A+\mu cot\beta) &   &  &   &  m^{ 2}_{ {U^{ c}_{ 3}}}+m^{ 2}_{ t}\\
&   &    &   &\quad\quad+{M^{ 2}_{ Z}}(cos2\beta)
                      ({\frac {2}{3}}sin^{ 2}{\theta_{ W}})
\end{array}\right)
\end{equation}
%
and the diagonalizing matrix is defined as
${\bf K}^{\tilde t} {\bf {\cal M}}^{ 2}_{ {\tilde t}}
{{\bf K}^{\tilde t}}^{\bf T}
=diagonal(m^2_{{\tilde t}_1},m^2_{{\tilde t}_2})$.
In the absence of \, SUSY breaking effects $m_{{\tilde f},{\tilde f'}} =
m_{f,f'}$ and ${\bf K}^{{\tilde f},{\tilde f'}}$ become the unit matrices. In
that limit $\Delta Q_V$ given above cancels against the corresponding
fermionic contribution as it should.
\vspace{1cm}
\subsection*{Neutralinos--Charginos ($\tilde{Z}$, $\tilde{C}$)}
The neutralino and chargino sector is perhaps the most awkward sector to deal
with owing to mixings originating from the electroweak symmetry breaking
effects. Their contributions are read from the graphs shown in Figure 2c,d.
In the following we shall denote by $\tilde{C}_i$ the two chargino states
(Dirac fermions) and  by
$\tilde{Z}_\alpha$  the four neutralinos states (Majorana fermions).
Recall that they are eigenstates of the following mass matrices
$^{\cite{spa}}$,
\begin{equation}
{\bf {\cal M}_{\tilde{C}}}=\left(  \begin{array}{cc}
M_2      &     -g\,v_2   \\
-g\,v_1   &      \mu
\end{array}  \right)
\end{equation}
\vspace{0.2cm}
\begin{equation}
{\bf {\cal M}_{\tilde{N}}}=\left( \begin{array}{cccc}
    M_1               &     0         &g^\prime v_1 / \sqrt 2
                                                 &  -g^\prime v_2/\sqrt 2 \\
    0                 &    M_2        &  -g v_1/\sqrt 2
                                                 &   g v_2/\sqrt 2       \\
g^\prime v_1/\sqrt 2  & -g v_1/\sqrt 2&       0
                                                 &        -\mu           \\
-g^\prime v_2/\sqrt 2  & g v_2/\sqrt 2 &      -\mu
                                                 &        0
\end{array}\right)
\end{equation}
where $v_1/\sqrt{2}$, $v_2/\sqrt{2}$ are the v.e.v.'s of the neutral
components of the Higgs field $H_1$ and $H_2$ respectively. If the ${\bf U}$,
${\bf V}$ matrices diagonalize ${\bf {\cal M}_{\tilde{C}}}$, i.e.
${\bf U{\cal M}_{\tilde{C}} V^{\dagger}}=diagonal$, and ${\bf O}$
(real orthogonal) diagonalizes the real symmetric neutralino mass matrix of
Eq.(19), ${\bf O^{T}{\cal M}_{\tilde{N}} O}=diagonal$, then the
electromagnetic and weak currents are given by
\begin{eqnarray}
{J_{em}^\mu}&=&{\sum_{i}} {\bar{\tilde C}}_i  \gamma^\mu   {\tilde C}_i \\
{J^\mu_+}&=&{ \sum_{\alpha,i}}{\bar {\tilde Z}}_\alpha \gamma^\mu
      ( P_R C^R_{\alpha i}+P_L C^L_{\alpha i}) {\tilde C}_i \\
{J^\mu_0}&=&{ \sum_{i,j}}{\bar {\tilde C}}_i \gamma^\mu
      ( P_R A^R_{ij}+P_L A^L_{ij}) {\tilde C}_j+
 {1 \over 2} { \sum_{\alpha,\beta}}{\bar {\tilde Z}}_\alpha \gamma^\mu
      ( P_R B^R_{\alpha \beta}+P_L B^L_{\alpha \beta}) {\tilde Z}_\beta
\end{eqnarray}
where
\begin{eqnarray}
C^R_{\alpha i}&=&-{\frac {1}{\sqrt 2}} O_{3 \alpha} U^{*}_{i2}-
                                     O_{2 \alpha} U^{*}_{i1} \;\;,\;\;
C^L_{\alpha i}=+{\frac {1}{\sqrt 2}} O_{4 \alpha} V^{*}_{i2}-
                                     O_{2 \alpha} V^{*}_{i1} \\
A_{ij}^h&=&[\cos^2\theta_W \delta_{ij}-{1 \over 2}(V_{i2}V_{j2}^* \delta_{hL}
       +U_{i2}^*U_{j2} \delta_{hR})]\;\;,\;\;h=L,R \\
B_{\alpha \beta}^L&=&{1 \over 2}(O_{3 \alpha}O_{3\beta}-O_{4 \alpha}O_{4\beta})
      \quad,\quad B_{\alpha \beta}^R=-B_{\alpha \beta}^L\:.
\end{eqnarray}
$P_{R,L}$ are the right/left handed projection operator $(1\pm \gamma_5)/2$.
The contributions of this sector to $\Delta k_{\gamma}$, $\Delta Q_{\gamma}$,
as calculated from the graph shown in Figure 2c, is as follows:
\begin{eqnarray}
\Delta k_{\gamma}&=&-\sum_{i, \al}\olocl \{ F_{\al i}
          [t^4+(R_{\al}-R_i-1)t^3+(2R_i-R_{\al})t^2    \nonumber\\
     & & + \usualp ] + G_{\al i} {{m_i m_\al}\over {M_W^2}}(4t^2-2t) \}
     {1 \over L_{\tilde Z}^2} \\
\Delta Q_{\gamma}&=&-8\sum_{i ,\al} F_{\al i} \olocl {t^3(1-t)\al
          (1-\al) \over  L_{\tilde Z}^2}
\end{eqnarray}

\begin{eqnarray}
L_{\tilde Z}^2&=&t^2+(R_i-R_{\al}-1)t+R_{\al}-\usual \\
F_{\al i}&=&\mid C_{\al i}^R \mid ^2+ \mid C_{\al i}^L \mid ^2 \quad,
  \quad G_{\al i}\;=\;(C_{\al i}^L C_{\al i}^{R*}+h.c.) \\
 & & R_{\al,i}\equiv{m_{\al,i}^2 \over M_W^2} \nonumber\
\end{eqnarray}
where the index $\al=1,2,3,4$ is the neutralino index and $i=1,2$ is the
chargino index. Note that we have not commited ourselves to a particular sign
convention for the masses $m_i$, $m_{\al}$ appearing to the sum in
the equation above for the $\Delta k_{\gamma}$.
Chiral rotations that makes these masses positive it also affects
the rotation matrices and should be taken into account.
\par
For the couplings $\Delta k_Z$, $\Delta Q_Z$ we get,\\
Graph of  Figure 2c :
\begin{eqnarray}
\Delta k_Z&=&-R\;\sum_{i,j,\al}\olocl \{ S^L_{ij\al} [t^2(1-t)(-t
     +R_i\al +R_j (1-\al)-R_{\al})\nonumber\\
   & &+\usualp]+ \,{m_im_j \over M_W^2} S^R_{ij\al} t^2 \nonumber \\
 & &-{m_im_{\al} \over M_W^2} (T^L_{ij\al}+T^R_{ij\al})[2t^2\al +t^2-t] \}
 {1 \over L^2_{ij\al}} \\
\Delta Q_Z&=&-8R\;\sum_{i,j,\al}
                  S^L_{ij\al}\olocl{t^3(1-t)\al (1-\al) \over L^2_{ij\al}}
\end{eqnarray}
where
\begin{eqnarray}
S^{L(R)}_{ij\al}&\equiv&(C^{L*}_{\al i}C^{L}_{\al j}A^{L(R)}_{ji}
                     +(L\rightleftharpoons R))\quad,\quad
T^{L(R)}_{ij\al}\equiv(C^{L*}_{\al i}C^{R}_{\al j}A^{L(R)}_{ji}
                     +(L\rightleftharpoons R)) \nonumber \\
L^2_{ij\al}&=&\al tR_i+t(1-\al)R_j+R_{\al}(1-t)-t(1-t)-\usual \nonumber  \\
R_{i,j,\al}&\equiv&{m_{i,j,\al}^2 \over M_W^2}\quad,\quad i,j=chargino
\quad indices \quad,\quad \al=neutralino \quad index
\nonumber\
\end{eqnarray}
\vspace{.5cm}
\\
Graph of Figure 2d :  \\
Same as in previous graph with $\{ i, j, \alpha \}$ replaced by
$\{ \rho, \sigma, i \}$  and
$ S^{L(R)}_{ij\rho}, T^{L(R)}_{ij\rho}$,
$L^2_{ij\rho}$ replaced by the following expressions: \\
\begin{eqnarray}
S'^{L(R)}_{\rho \sigma i}&\equiv&-(C^{L*}_{\rho i}C^{L}_{\sigma i}
            B^{L(R)}_{\rho \sigma}+(L\rightleftharpoons R)) \nonumber \\
T'^{L(R)}_{\rho \sigma i}&\equiv&-(C^{R*}_{\rho i}C^{L}_{\sigma i}
            B^{L(R)}_{\rho \sigma}+(L\rightleftharpoons R))   \nonumber \\
L'^{L(R)}_{\rho \sigma i}&=&\al tR_{\rho}+t(1-\al)R_{\sigma}
         +R_i(1-t)-t(1-t)-\usual \nonumber \\
{\sigma,\rho}&=&neutralino \quad indices \quad,\quad i=chargino \quad index
\nonumber
\end{eqnarray}
\vspace{1cm}
\subsection*{Higgses ($H_0,h_0,A,H_{\pm}$)}
There are five physical Higgs bosons which survive electroweak symmetry
breaking. Two of these, $H_0$ and $h_0$, are neutral and $CP$ even, while
a third $A$, is neutral and $CP$ odd. The remaining Higgses bosons,
$H_{\pm}$, are charged. At the tree level the lightest of these, namely
$h_0$, is lighter than the $Z$ gauge boson itself. However
it is well known that radiative corrections which are
due to the heavy top are quite large and should be taken into account. These
modify its tree level mass by large amounts
$\delta m_{h_0}^2\sim g^2{(m_t^4 / M_W^2)}\ln({m_{\tilde t_1}^2
m_{\tilde t_2}^2 /m_t^4})$ which may push its mass up  to
values exceeding $M_Z$ ; in some cases up to $\simeq 130 \: GeV$.
$h_0$ turns out to yield the largest contributions of all Higgses
to the TGV's since the remaining Higgses have
large masses of the order of the SUSY breaking scale.
At the tree level the masses of all Higgs bosons involved are given by the
following expressions :
\begin{eqnarray}
{m^2_A}&=& -{2m_3^2 \over \sin2\beta}\quad,\quad (m_3^2\equiv B\mu)   \\
{m_{ H_0,h_0 }^2}&=&{\frac {1}{2}} \{  {(m^2_A+M_Z^2)}^2\pm
  \sqrt {   {(m^2_A+M_Z^2)}^2 -4{M_Z^2} {m^2_A} {cos^2}(2\beta) }
                                                                 \}  \\
m^2_{H_\pm}&=&m^2_A+ M_W^2
\end{eqnarray}
The Higgs contributions can
be expressed in terms of their masses and an angle $\theta$, which relates
the states $S_1 \equiv \cos \beta \: (Real \; H_1^0) +
\sin \beta \: (Real \; H_2^0)$\, , \,
$S_2 \equiv -\sin \beta \: (Real \; H_1^0) +
\cos \beta \: (Real \; H_2^0)$ to the mass eigenstates $h_0,H_0$.
The state $S_1$ is the SM Higgs boson which however is not a mass
eigenstate since it mixes with $S_2$.
When $\sin ^2 \theta =1$ such a mixing does not occur and $h_0$ becomes
the standard model Higgs bosons $S_1$.
\par
The contributions of the Higgs bosons to
$\Delta k_{\gamma}$, $\Delta Q_{\gamma}$, follow from the graphs shown in
Figure 3 and are as follows,
\\
\vspace{.5cm}
Graphs of Figures 3a and 3b :
\begin{eqnarray}
&&A\;:\qquad \Delta k_{\gamma}=D_2(R_A,R_+) \quad , \quad
            \Delta Q_{\gamma}=Q(R_A,R_+) \\
&&h_0\,:\qquad \Delta k_{\gamma}\:=\: \sin^2 \theta \: D_1\,
                         (R_{h_0})\:+\:\cos^2 \theta \: D_2(R_{h_0},R_+)  \\
&&\quad \qquad \hspace*{4.5mm} \Delta Q_{\gamma}\:=\: \sin^2 \theta \:
                             Q(R_{h_0},1)\:+\: \cos^2 \theta
                                             \:Q(R_{h_0},R_+)\\
&&H_0\,:\quad \, \; \; As \,\, in\quad h_0\quad with
                \quad R_{h_0} \rightarrow R_{H_0}
\quad and
       \quad  \sin^2 \theta \rightleftharpoons  \cos^2 \theta   \\
&& \nonumber\\
&&  R_a\equiv{(m_a/M_W)}^2\;,\; a=h_0,H_0,A,H_{\pm} \;\;,\;\;
    \sin ^2\theta={M_A^2+M_Z^2\sin ^2 2\beta -M_{h_0}^2 \over
            M_{H_0}^2-M_{h_0}^2}\: . \nonumber\
\end{eqnarray}
The functions $D_{1,2}, Q$ appearing above are defined in Appendix A.
\par
For the $\Delta k_Z$, $\Delta Q_Z$ form factors there are more graph
contributing. These are shown in Figures 3c--3f.
{}From the graphs shown in Figure 3 we pick the following contributions,
\\
\vspace{.5cm}
Graph of Figure 3a:
\begin{eqnarray}
\Delta k_{Z}&=&{1\over 4}\{ (\cos^2 \theta) \olocl\; [(4-2R)t^4+(R-2)
     (R_{H_0}+2)t^3 \nonumber\\
   & &+(2R_{H_0}-RR_{H_0}+8-2R)t^2] {1 \over L^2_{H_0}}
     + (\sin^2 \theta)\times (H_0 \rightarrow h_0) \} \\
\Delta Q_{Z}&=&(2-R) \{ (\cos^2 \theta) \olocl {t^3(1-t) \al (1-\al)
   \over L^2_{H_0} } \nonumber\\
 & & +(\sin^2 \theta)\times (H_0 \rightarrow h_0) \} \\
L^2_{H_0}&\equiv &t^2+R_{H_0}(1-t)-\usual \quad , \quad
R_{H_0,h_0} \equiv  {m_{H_0,h_0}^2 \over M_W^2} \nonumber
\end{eqnarray}
\vspace{.5cm}
\\
Graph of Figure 3b :
\begin{eqnarray}
&& \Delta k_{Z}=({2-R \over 2}) \{ D_2(R_A,R_{+})+\sin^2 \theta
     \;D_2(R_{H_0},R_{+})+\cos^2 \theta \;D_2(R_{h_0},R_{+}) \} \\
&& \Delta Q_{Z}=({2-R \over 2}) \{ Q(R_A,R_{+})+\sin^2 \theta\;
    Q(R_{H_0},R_{+}) +\cos^2 \theta \;\;Q(R_{h_0},R_{+}) \}
\end{eqnarray}
\vspace{.5cm}
Graphs of Figures 3c and 3d :
\begin{eqnarray}
\Delta k_{Z}&=&{R\over 2} \{ (\sin^2 \theta) \olocl\; {t^2(1-t)(1+R_{+}
     -R_A\al -R_{H_0}(1-\al)-2t) \over {\tilde L}^2_{H_0}} \nonumber\\
& & \hspace{7.2cm} + (\cos^2 \theta)\times (H_0 \rightarrow h_0) \} \\
\Delta Q_{Z}&=&2R \{ (\sin^2 \theta) \olocl {\al (1-\al) t^3(1-t)
   \over {\tilde L}^2_{H_0} }
    +(\cos^2 \theta)\times (H_0 \rightarrow h_0) \} \\
{\tilde L}^2_{H_0}&\equiv &-t(1-t)+R_A\al t+R_{H_0}(1-\al)t
                   + R_{+}(1-t)-\usual
\end{eqnarray}
\\
and finally, \\
Graphs of Figures 3e and 3f :
\begin{eqnarray}
\Delta k_{Z}&=&{R\over 2}\{ (\cos^2 \theta) \olocl\;
    [-6\al t^2+(t^3-t^2)  (2(t-1)+(R-R_{H_0})\al +R_{H_0}) \nonumber\\
   & &+2(R-1)\al t^2] {1 \over {\hat L}^2_{H_0}}
     + (\sin^2 \theta)\times (H_0 \rightarrow h_0) \} \\
\Delta Q_{Z}&=&2R \{ (\cos^2 \theta) \olocl {\al (1-\al)t^3 (1-t)
   \over {\hat L}^2_{H_0} } \}
    +(\sin^2 \theta)\times (H_0 \rightarrow h_0) \} \\
{\hat L}^2_{H_0}&\equiv &(1-t)^2+R\al t+R_{H_0}t(1-\al)-\usual
\end{eqnarray}
In the most of the parameter space
the Higgses $A, H_{\pm}$ and $H_0$ turn out to be rather heavy having
masses of the order of the SUSY breaking scale; therefore all graphs in
which at least
one of these participates are small. At the same time $\sin^2 \theta$ has a
value very close to unity. Thus the dominant Higgs contribution arises solely
from the graphs of Figures 3a,e and 3d in which a $h_0$ is exchanged. This
is exactly what one gets in the SM with $h_0$ playing the role of the SM
Higgs boson.
%
\newpage
\section*{5. The Absortive Parts of the TGV's}
\par
The contributions of the TGV's presented so far
have also imaginary (absortive) {\mbox{parts}}
which show up as soon as one passes the
thresholds associated with the particles exchanged in the one loops graphs.
Since the majority of the one loop expressions encountered have a triangle
structure some of these thresholds can be anomalous and this depends on
the masses of the particles circulating in the loop.
\par
The absortive parts can be readily calculated using the $i \epsilon$
prescription. Actually the denominators of the Feynman integrals involved
carry a small positive imaginary part and it is a matter of a proper
algebraic manipulation to pick up the relevant imaginary parts of all
integral expressions presented in the previous sections.
All graphs yielding nonvanishing contributions to
${\Delta{ k_{\gamma ,Z}}  },{\Delta{ Q_{\gamma ,Z}}}$ have a
triangular structure with the exception of some of the gauge boson graphs
which involve the quartic gauge boson coupling. For these
we have   ${ \Delta{ Q_{\gamma ,Z}}}=0$ with  ${\Delta{ k_{\gamma ,Z}}}$
given by $^{\cite{argy}}$ \\
\begin{eqnarray}
   \Delta k_{\gamma,Z}&=&{21 \over {2}}{\sin^2}{\theta_W}
                - {3 \over {2R}}  \int_{0}^{1}dt
                { {2t^3-(8+R)t^2+4Rt}  \over
                { t^2+R(1-t)+i\epsilon  }} \nonumber  \\
              &+&   {9 \over 2}
         {  \int_{0}^{1}dt \; {\ln (1-\usual)}  }
\end{eqnarray}
where  $(R\equiv M_Z^2/M_W^2)$. The first of the integrals in the expression
above does not have any discontinuity
since the denominator never vanishes
{\footnote{
Actually this integral arises from a diagram that has the structure of a two
point Greens function. It involves the quartic gauge coupling where one of its
legs is the  $\gamma$ or $Z$ and the other is one of the on shell
external $W$'s. Since the $W$'s are on their mass shell it does not depend on
$Q^2$. }}
The second integral developes an absortive part when
the argument of the logarithm becomes negative. Therefore it has an Imaginary
part given by \\
\begin{eqnarray}
Im \, {\Delta{k_{\gamma,Z}}} \, = \,
{{9 \pi} \over {2}} {\Theta (Q^2-M^2_W)} {\sqrt { (1-{{M^2_W} \over {Q^2}})} }
\end{eqnarray}

For the pinch contributions we have an absortive part arising from the single
$dt$ integrations appearing in the Eqs. (13),(14) and an additional
contribution which stems from the double $dt, \,  d\al$ integrations of
these equations. The later yield absortive parts, denoted by
$\Delta_{\gamma,Z}$, which have a structure akin to those of the
triangle graphs. The former yield absortive parts which are easily calculated
leading to the following results: \\
\begin{eqnarray}
Im \, { {\hat \Delta} k_{\gamma } }   &=&
   \pi \,  {\sin^2}{\theta_W} \, {{\Theta (Q^2-M^2_W)} \over
  { \sqrt {(1-{{M^2_W} \over {Q^2}} )}}     }
   + {\Delta_{\gamma}}  \\
Im \, { {\hat \Delta} k_{Z } }   &=&
   \pi \, {{\sin^2}{\theta_W}} \, (1-R \, {{M_W^2} \over{4 {Q^2}}}) \,
   {{\Theta (Q^2-M^2_W)} \over {\sqrt {(1-{{M^2_W} \over {Q^2}})}}  }
   + {\Delta_Z}
\end{eqnarray}

The absortive parts of all triangle graphs as well as the contributions
$\Delta_{\gamma,Z}$ in the Equations (51) and (52) above can be inferred by
calculating the imaginary part of the  integral  \\
\begin{eqnarray}
I \, \equiv \, {\olocl {{P_2 {\alpha}^2 +P_1 \alpha +P_1} \over
  { \rho +  \sigma \alpha + \usual }} }
  \label{inte}
\end{eqnarray}
where $P_{0,1,2}$ as well as $\rho,\sigma$ are functions of the variable $t$
alone. The absortive part of this integral is not difficult to calculate, and
the details are presented in the Appendix B. In its final form it can be
expressed as an single integral over the variable $t$
(see Eq. (58), Appendix B)  which can be
integrated numerically using special numerical routines. Having expressed
the Imaginary parts of all contributions involved, as integrals in $t$ of
known functions of $t$ and the energy variable  $Q^2$, we are ready to
proceed to numerical computations.
\par
The strategy we follow for the evaluation of the absortive parts will also
apply to the real (dispersive) parts of the form factors under consideration.
In fact wherenever a double $\olocl$ integrations are encountered we first
perform the $\int_{0}^{1} d\alpha$ integrations explicitly and subsequently
evaluate the $\int_{0}^{1} dt$ integrations numerically for various values
of $Q^2$ and the MSSM parameters. The details of the numerical analysis is the
subject of the following section.
\vspace{1cm}
\section*{6. Numerical Analysis -- Conclusions }

As discussed in the previous section both dispersive and absortive parts of
the trilinear $WW\gamma$, $WWZ$ vertices can be cast as single integrals of
known functions of $t$ and $Q^2$, which also depend on the physical masses
of all particles involved.
These integrations we have numerically carried out using special
routines of the  FORTRAN Library $IMSL$ available to us.
The advantage of using
this facility is that it leads to reliable results even in cases where the
integrands exhibit fast growth at some points or have a rapid oscillatory
behaviour. The inputs in these calculations are the value of the
energy variable $Q^2$ and the arbitrary parameters of the
MSSM discussed in the section 2.
\par
In our numerical analysis we have taken all Yukawa couplings, but that of
the top quark vanishing, which is a very good approximation especially for
the reason that the trilinear gauge boson vertices under consideration are
already of one loop order. This approximation however holds provided the
value of the parameter $\tan \beta (M_Z)$ which sets the relative strentgth
of the v.e.v's of the two Higgses involved are not large  $\le 10$.  For
larger values the bottom Yukawa coupling should be also considered in the
RGE's of all running parameters involved.
However this approximation little affects our numerical results
for the form factors under consideration.
\par
With the experimental inputs $M_Z=91.18 \: GeV$, $\sin^{2} {\theta_W} =.239$,
$\alpha_{em}(M_Z)=1/129$ and  $\alpha_{s}(M_Z)=.117$ and with given values for
the arbitrary parameters  $\tan \beta (M_Z)$, $m_t (M_Z)$,
$A_0$, $m_0$, $M_{1/2}$
we run our numerical routines in order to know the mass spectrum and the
relevant mixing parameters necessary for the evaluation of the form factors
given in the
previous sections. Throughout the analysis we have taken
{\footnote { Our results for
${\Delta k}_{V}$, ${\Delta Q}_{V}$ are not sensitive to the choice of the
angle $\beta$.}} $\tan \beta (M_Z) \le 10$,
but its value cannot be taken arbitrarilly small. Actually given
$m_t (M_Z)$ the parameter $\tan \beta (M_Z)$ is forced to a minimum value
otherise Landau poles are encountered making the top Yukawa coupling
${{h_t^2}/ { (4 \pi)^2}}$ getting values $\geq 1$ outside the validity of
the perturbative regime.
\par
For the running top quark mass $m_t(M_{ Z})$   we took values
in the whole range from $130\: GeV$ to $190\: GeV$,
although small values of $m_t$ are already ruled out experimentally
in view of the recent $CDF$ and $D0$ results which both quote a mass for
the top quark larger than about $170\: GeV$ $^{\cite{abe}}$.
The physical top quark masses emerging out are slightly larger by
about $3 \% $
{ \footnote{ The physical top quark mass $M_t$ in the DR scheme is given
by $M_t=m_t(M_t)/(1+{{ 5\alpha_3(M_t)}/{3\pi}}) $. }  }.

As for the soft SUSY breaking parameters
$A_0,m_0,M_{1/2}$ we scan the three dimensional parameter space from
$\simeq 100\: GeV$ to $1\: TeV$.
This parameter space can be divided into three
main regions: \\ \\
i) $A_0 \simeq m_0 \simeq M_{1/2} $ (all SUSY breaking terms comparable)
\newline
ii) $A_0 \simeq m_0 \ll M_{1/2} $ (the gaugino mass is the dominant source
of SUSY breaking ) \newline
iii) $M_{1/2} \ll  A_0 \simeq m_0 $ ($A_0, m_0$ dominate over $M_{1/2}$) \\ \\
Case ii) covers the no-scale models for which in most of the cases the
preferable values are $A_0=m_0=0$ while case iii) the light gluino case.
\par
Regarding the values scanned for the energy variable $Q^2$ we moved both in
the timelike and spacelike region for values ranging from
$\mid Q^2 \mid \, = \, 0$
to $\sqrt{ \mid Q^2 \mid} \,=\,10^5 M_W$. For the timelike case, which is of
relevance for future collider experiments, this corresponds to values of
$\sqrt s$ ranging from $0\: GeV$ to about $600\: M_W$. For comparison we
quote that $\sqrt s$ at LEP200 will be $190\: GeV$ that is it just exceeds
the two $W$'s production threshold energy $2 M_W$. Both in the spacelike
and timelike energy region as soon as $\sqrt s$  exceeds $\simeq$ few $TeV$
the contributions of each sector separately becomes negligible approaching
zero as the energy increases in accord with unitarity requirements.
\par
Sample results are presented in Tables I and II for values of
$(A_0,m_0,M_{1/2})$ equal
to  $(300,$ $300,$ $300)$, $(0,0,300)$ and $(300$, $300$, $80)$ $GeV$
representative of the cases i),ii) and iii) respectively discussed above.
The inputs for
the remaining parameters are $tan\beta (M_Z) = 2$, $m_t(M_Z)=170\: GeV$.
The value of $\sqrt s$ in these tables are respectively $190$ and $500\:GeV$,
corresponding to the center of mass energy of LEP200 and NLC. In the same
tables for comparison we give the SM predictions for Standard Model Higgs
masses $50, 100$ and $300\: GeV$. With the inputs given above the typical
SUSY breaking scale lies in somewhere between $2 M_W$ and $.5 \:TeV$.
Although many
sparticle thresholds exist in this region, as for instance the lightest
of the sleptons and squarks as well as the lightest of the neutralinos
and charginos, especially when $M_{1/2}$ is light, these thresholds do not
result in any enhancement of the form factors
$\Delta k_{\gamma,Z}$, $\Delta Q_{\gamma,Z}$. Increasing the value of the
dominant SUSY breaking scale the supersymmetric contributions to these
quantities become less important approaching zero values. Of all sectors
the sfermions yield the smaller contributions in the entire parameter space
even in cases where due to large electroweak mixings some of the squarks,
namely one of the stops, are relatively light. The supersymmetric Higgses
yield contributions comparable to those of the SM, provided the latter
involves a light Higgs with mass around the $100\:GeV$ scale. The bulk of the
Higgs contributions is due to the lightest $CP$ even neutral $h_0$.
As discussed in previous sections the
dominant Higgs contributions come from the diagrams of Figures 3a,e
and 3f in which a light Higgs $h_0$ is exchanged. These are actually
the only sources of Higgs contributions in the SM with $h_0$ replaced
by the Standard Model Higgs boson.
\par
The last sector
to be discussed is the neutralinos and charginos which in some cases,
depending on the given inputs, can accomodate light states. Their
contributions in that case are not necessarily small and is the principal
source of deviations from the SM predictions. The
contributions of the neutralinos and charginos are sensitive to the input
value for the soft gaugino mass $M_{1/2}$ being more important for values
$M_{1/2} < 100 \; GeV$. For such values of the soft gaugino mass and in the
unphysical region,
$0 < \sqrt{s} < 2 M_W $ they are enhanced, due
to the development of an anomalous threshold in this region, which
results to sizable corrections to the magnetic dipole and electric
quadrupole moments of the $W$ boson $^{\cite{spa,lahanas}}$.
\par
In Figure 4 and 5 we plot the contributions of the Higgses and of the
neutralino - chargino sector to
$\Delta k_{\gamma,Z}$, $\Delta Q_{\gamma,Z}$ for the most physically
interesting case $(300$, $300$, $80)$ $GeV$ and for values of $\sqrt{s}$
ranging from $0\: GeV$ to $1\: TeV$.
The region from $0$ to $2M_W$ is unphysical since the external $W$'s have
been taken on their mass shell. At $s = 0$ the quantities
$\Delta k_{\gamma}$, $\Delta Q_{\gamma}$ are linearly related to the magnetic
moment and electric quadrupole moments of the $W$-boson.
\par
The structure shown
in the Higgs contributions for $\sqrt{s} \le 200 \:GeV$ is  due to the
lightest of the Higgses. One observes a fast fall off as we increase the
energy to values above $\approx 200\: GeV$. In the neutralino and chargino
sector, and for the $\mu > 0$ case, a sharp peak is observed in the
unphysical region, $\sqrt{s} < 2 M_W$ due to the appearance of the
anomalous threshold discussed previously and the contributions of this sector
is
substantially enhanced.
However such an enhancement does not occur in the physical region since their
contributions fall rapidly to zero as we depart from the unphysical region
to values of energies above the two $W$ production threshold. This behaviour
is clearly seen in Figure 5. The structure observed at energies around
$700\: GeV$ comes from the graph of Figure (3d) and is due to the fact that
for these energies we are close to thresholds associated with the heavy
neutralino states.
\par
The total contributions to the TGV's both in the MSSM
and SM are shown in Figures 6 to 8. We display both dispersive and
absortive parts of the form factors under consideration.
One notices that all form factors
tend to zero fairly soon with increasing the energy
reaching their asymptotic values at energies $\sqrt{s} \approx few TeV$ in
agreement with unitarity constraints.
\par
Our conclusion is that for energies
$2 M_W < \sqrt{s} < 200 \; GeV $ the MSSM predictions differ in general
from those of the
SM but they are of the same order of magnitude. Deviations from the SM
predictions to be detectable require sensitivities reaching the per mille
level and hence  unlikely to be observed at LEP200. If deviations from the
SM predictions are observed at these energies will be the signal of new
underlying dynamics which however will not be of supersymmetric nature.
At higher energies  SM and
MSSM predictions fall rapidly to zero, due to unitarity,
getting smaller by almost an order of magnitude already at energies
$\sqrt{s} \approx .5 \; TeV $. As a result, the task of
observing deviations from the SM which are due to supersymmetry
becomes even harder at these energies demanding higher experimental
accuracies.

\vspace*{2cm}
\noindent
{\bf {\large {Acknowledgements}}}
\\
\\
*\vspace*{4mm}
Work supported by EEC Human Capital and Mobility Program, CHRX -- CT93 - 0319.
\newpage

\appendix
\section{Appendix}
\par
The functions $D_1(r), D_2(r,R)$ and $Q(r)$ through which the Higgs
contributions are expressed are defined as follows:
\begin{eqnarray}
&&   D_1(r)\equiv{1\over 2} \olocl
        { 2t^4+(-2-r)t^3+(4+r)\,t^2   \over
        t^2+r(1-t)-\usual }      \\
&&    D_2(r,R)\equiv{1\over 2} \olocl
        { 2t^4+(-3-r+R)t^3+(1+r-R)t^2   \over
        t^2+(-1-r+R)t+r-\usual }  \\
&&    Q(r,R)\equiv 2 \olocl
        { t^3(1-t)\al (1-\al) \over
         t^2+(-1-r+R)t+r-\usual }
\end{eqnarray}
\vspace{.5cm}
\section{Appendix}
\par
The Imaginary part of the integral I defined in the main text (see Eq.
\ref{inte} )  is given by \\
\begin{eqnarray}
Im \, I = -\pi \olocl ({P_2 {\alpha}^2 +P_1 \alpha +P_0}) \,\times
\nonumber \\
\delta \, ( \use + \sigma \alpha +\rho )
\end{eqnarray}

The roots $\rho_{1,2}$ of the argument of the delta function in the
expression above are given by \\

\begin{eqnarray}
\lambda_{1,2} \, = \, {1 \over 2} \, [(1- {\sigma \over {\hat{s} t^2} })
\pm \sqrt{ (1- {\sigma \over {\hat{s} t^2} })^2-4{\rho \over {\hat{s} t^2}}}]
\quad ,\quad    \hat s \equiv {{4 Q^2} \over {M_W^2}}
\nonumber
\end{eqnarray}
The imaginary part $Im \, I$ vanishes if \\
\begin{eqnarray}
(1- {\sigma \over {\hat{s} t^2} })^2-4{\rho \over {\hat{s} t^2}} < 0  .
\nonumber
\end{eqnarray}
After a straightforward calculation one arrives at the following result\\

\begin{eqnarray}
Im \, I \,=\, -\pi {\sum_{i=1,2}} {\int_{0}^{1} dt}
(P_2 \lambda_{i}^{2}+ P_1 \lambda_{i}+P_0) \,{\Theta (\lambda_i)}
{\Theta (1-\lambda_i)} \times \nonumber   \\
{{ \Theta((\hat s t^2-\sigma)^2-4\, \rho t^2 \hat s)}  \over
{ \sqrt { (\hat s t^2-\sigma)^2-4\, \rho t^2 \hat s }} }
\end{eqnarray}

\newpage

\newpage
\noindent
{\large {\bf Table Captions}}

\vspace{.7cm}
\noindent
{\bf Table I}:\quad MSSM predictions for $\Delta k_{\gamma, Z}$,
$\Delta Q_{\gamma, Z}$, in units of $g^2/{16\pi^{2}}$,
for three different inputs of $A_0, m_0, M_{1/2}$.
Both $\mu >0$ and $\mu <0$ cases are displayed.
The energy is ${\sqrt{s}} = 190\: GeV$. The SM  predictions for Higgs masses
$50, 100$  and $300$ $GeV$ respectively are also displayed.

\vspace*{.5cm}
\noindent
{\bf Table II}:\quad MSSM predictions for $\Delta k_{\gamma, Z}$,
$\Delta Q_{\gamma, Z}$, in units of $g^2/{16\pi^{2}}$,
for three different inputs of $A_0, m_0, M_{1/2}$.
Both $\mu >0$ and $\mu <0$ cases are displayed.
The energy is ${\sqrt{s}} = 500 \:GeV$. The SM  predictions for Higgs masses
$50, 100$  and $300$ $GeV$ respectively are also displayed.

\vspace*{1.8cm}
\noindent
{\large {\bf Figure Captions}}

\vspace{.7cm}
\noindent
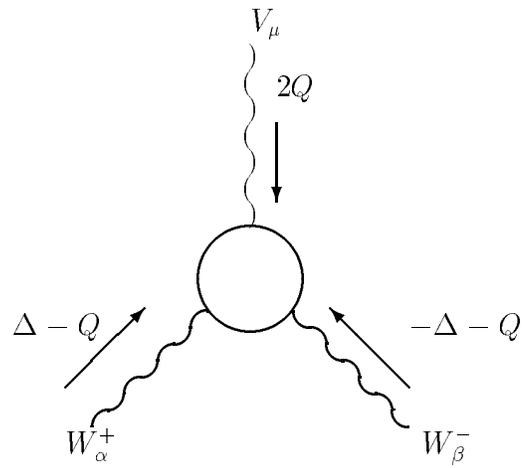
{\bf Figure 1}:\quad The kinematics of the $WWV$ vertex.

\vspace{.4cm}
\noindent
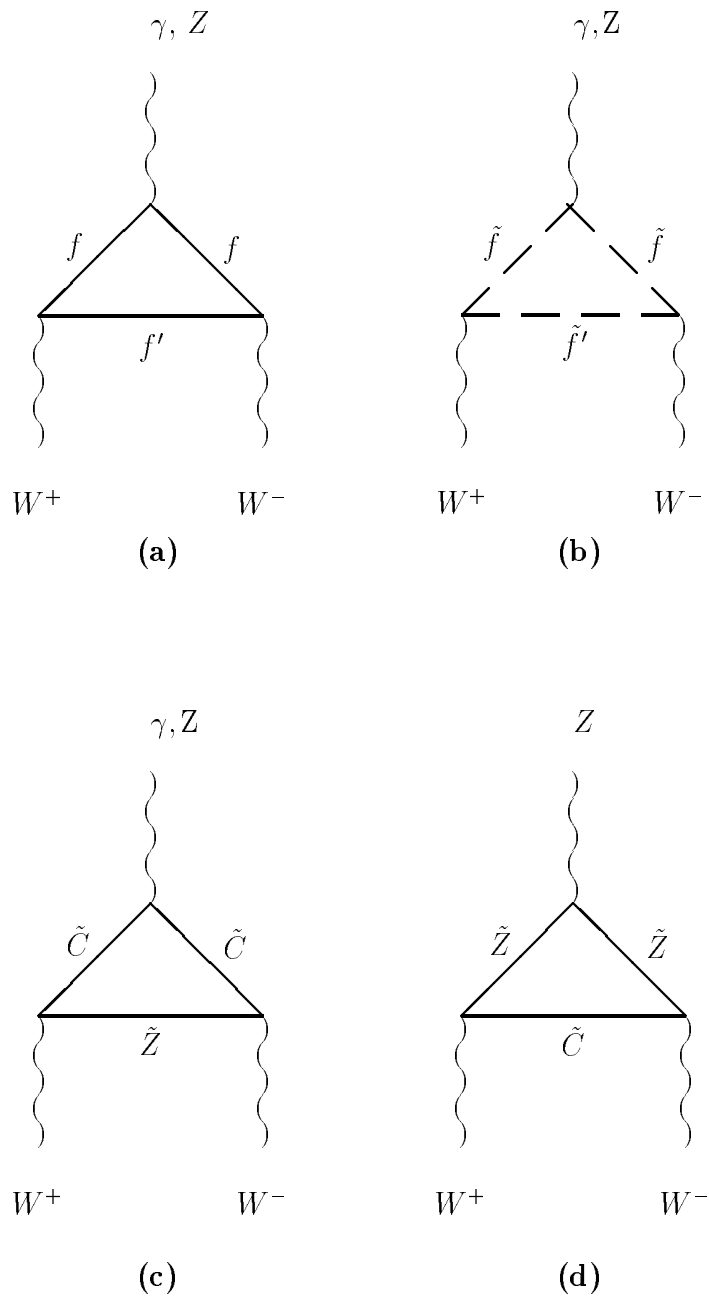
{\bf Figure 2}:\quad Fermion and sfermion contributions to
the $WWV$ vertices [figs. a,b]. Neutralino ($\tilde Z$) and
Chargino  ($\tilde C$) contributions to $WW\gamma$ [fig. c]
and $WWZ$ [figs. c,d] vertices.

\vspace{.4cm}
\noindent
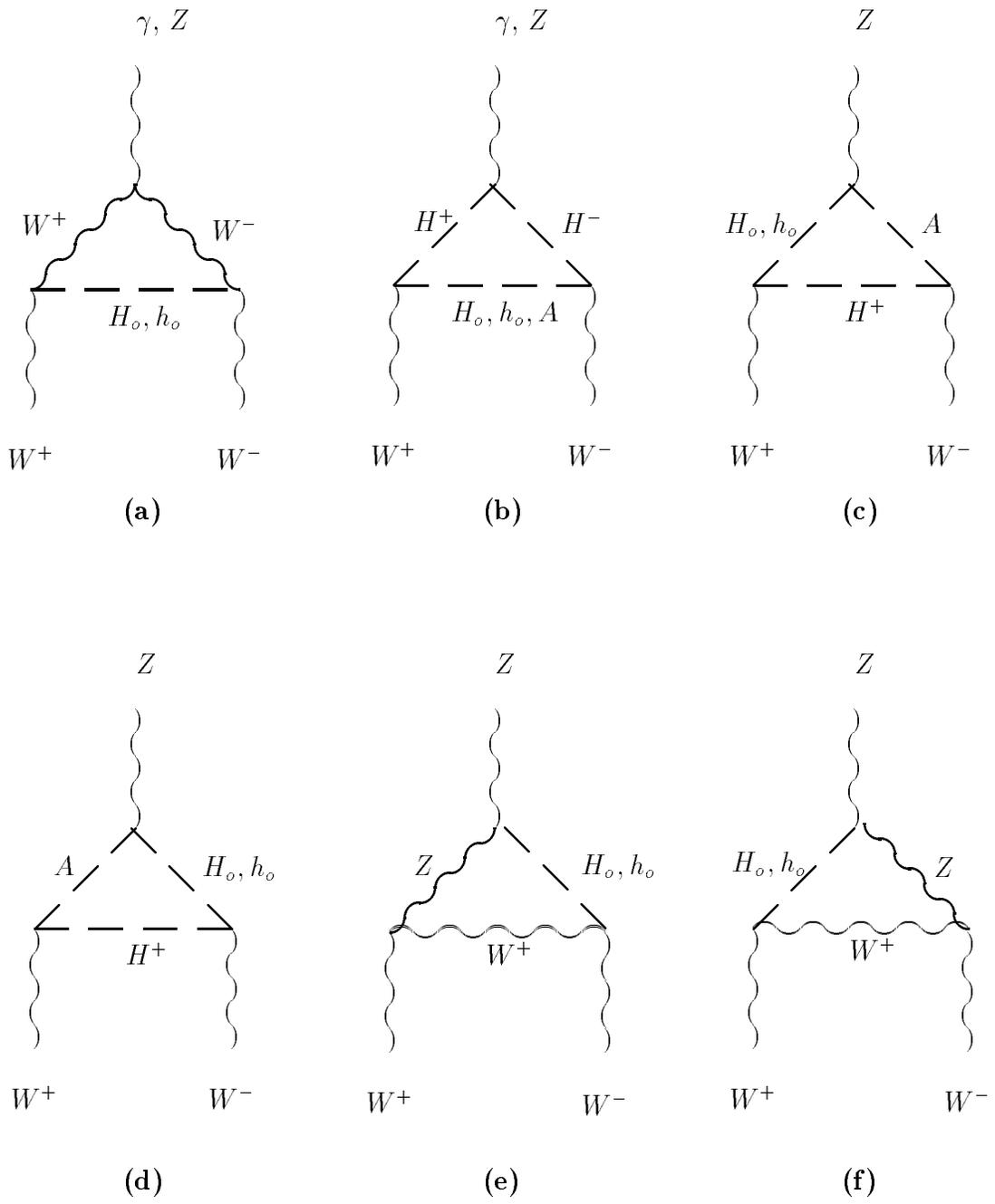
{\bf Figure 3}:\quad Higgs graphs contributing to the $WW\gamma$
and $WWZ$ trilinear vertices. $h_0$ denotes the lightest neutral Higgs boson.

\vspace{.4cm}
\noindent
{\bf Figure 4}:\quad MSSM Higgs contributions to the dispersive parts of
${\Delta k}_{\gamma}$, ${\Delta Q}_{\gamma}$  (solid lines) and
${\Delta k}_{Z}$, ${\Delta Q}_{Z}$ (dashed lines), in units of
$g^2/{16\pi^{2}}$, as functions of the energy $\sqrt s$. The inputs are
$(A_0, m_0, M_{1/2})$ $\,=\,$ $(300, 300, 80)$ $GeV$, $\tan \beta \, = \, 2$,
$m_t \, =\,170 \, GeV$. Both $\mu > 0$ and $\mu < 0$ cases are displayed.
The vertical dotted line indicates the position of the $2W$ production
threshold.

\vspace{.4cm}
\noindent
{\bf Figure 5}:\quad As in Figure 4 for the Neutralino and Chargino
contributions.

\vspace{.4cm}
\noindent
{\bf Figure 6}:\quad MSSM predictions for ${\Delta k}_{\gamma,Z}$
${\Delta Q}_{\gamma,Z}$. The parameters are as in Figure 4.

\vspace{.4cm}
\noindent
{\bf Figure 7}:\quad MSSM predictions for the absortive parts of
${\Delta k}_{\gamma,Z}$, ${\Delta Q}_{\gamma,Z}$ . The parameters are as
in Figure 4.

\vspace{.4cm}
\noindent
{\bf Figure 8}:\quad SM predictions for the dispersive [figs. a,b]
and absortive [figs. c,d] parts
of ${\Delta k}_{\gamma,Z}$, ${\Delta Q}_{\gamma,Z}$ for a Standard Model
Higgs mass equal to $100 \, GeV$ and $m_t \, =\,170 \, GeV$.
%
\newpage
\begin{table}
\begin{center}
\vspace*{1.cm}
\begin{tabular}{|c|ccc|} \hline
\multicolumn{4}{|c|}{ \bf TABLE I} \\
\hline
$A_0, m_0, M_{1/2}$ &300, 300, 80&[300, 300, 300]&(0,0,300)\\
 & $\tan \beta =2$ & $m_t= 170 \:GeV$ &  \\
\hline
$ \sqrt{s} = 190 \:GeV$  & $\mu>0$ & & $\mu<0$ \\
\hline
$\Delta k_{\gamma}$ & -1.938 [-1.742] (-1.762)&  & -1.733 [-1.767] (-1.791)\\
$\Delta Q_{\gamma}$ &  0.906 [ 0.526] ( 0.538)&  &  0.300 [ 0.529] ( 0.540) \\
$\Delta k_{Z}$      & -2.282 [-2.155] (-2.154)&  & -2.016 [-2.146] (-2.140) \\
$\Delta Q_{Z}$      & -0.354 [ 0.345] ( 0.334)&  & -1.193 [ 0.363] ( 0.357) \\
\hline
 $SM$   &$\Delta k_{\gamma} =$ -2.005, -1.735, -2.118 &
       &$\Delta k_{Z} =$ -1.350, -2.437, -1.404 \\
$predictions$ &$\Delta Q_{\gamma} =$ .524, .530, .503   &
        &$\Delta Q_{Z} =$ .507, .533, .481  \\
\hline
\end{tabular}
\label{tab1}
\end{center}
\end{table}
\vspace*{2.5cm}
\begin{table}
\begin{center}
\begin{tabular}{|c|ccc|} \hline
\multicolumn{4}{|c|}{ \bf TABLE II} \\
\hline
$A_0, m_0, M_{1/2}$ &300, 300, 80&[300, 300, 300]&(0,0,300)\\
 & $\tan \beta =2$ & $m_t= 170\: GeV$ &  \\
\hline
$ \sqrt{s} = 500 \:GeV$  & $\mu>0$ & & $\mu<0$ \\
\hline
$\Delta k_{\gamma}$ & -.262 [-.151] (-.191)&  & -.310 [-.207] (-.259) \\
$\Delta Q_{\gamma}$ & 0.150 [0.030] (0.056)&  & 0.146 [0.041] (0.069) \\
$\Delta k_{Z}$      & 0.121 [0.204] (0.209)&  & 0.240 [0.191] (0.198) \\
$\Delta Q_{Z}$      & 0.358 [-.407] (-.427)&  & 0.256 [-.325] (-.352) \\
\hline
 $SM$   &$\Delta k_{\gamma} =$-.250, -.168, .046 & \hspace*{4.5cm}
       &$\Delta k_{Z} =$ .147, .208, -.036\\
$predictions$ &$\Delta Q_{\gamma} =$ .054, .057, .064   &
        &$\Delta Q_{Z} =$ .057, .058, .077  \\
\hline
\end{tabular}
\label{tab2}
\end{center}
\end{table}
\vspace*{1.5cm}
\end{document}